# Ion energy distribution function in very high frequency capacitive discharges excited by saw-tooth waveform

**Sarveshwar Sharma[1,2], Nishant Sirse[3], Animesh Kuley[4], and Miles M Turner[5]**

[1]Institute for Plasma Research, Gandhinagar-382428, India
[2]Homi Bhabha National Institute, Anushaktinagar, Mumbai-400094, India
[3]Institute of Science and Research, IPS Academy, Indore-452012, India
[4]Department of Physics, Indian Institute of Science, Bangalore, India
[5]School of Physical Sciences and National Center for Plasma Science and Technology, Dublin City University, Dublin 9, Ireland

Email: nishantsirse@ipsacademy.org

**Abstract**

Tailoring *ion energy distribution function* (*IEDF*) is vital for advanced plasma processing applications. Capacitively coupled plasma (*CCP*) discharges excited using non-sinusoidal waveform have shown its capability to control *IEDF* through generation of *DC* self-bias. In this paper, we performed a particle-in-cell simulation study to investigate the *IEDF* in a symmetric capacitive discharge excited by saw-tooth like current waveform at a very high frequency (VHF).  At a constant driving frequency of 27.12 MHz, the simulation results predict that the ion energy symmetry scales with the discharge current amplitude and the *IEDF* turn into a bi-modal distribution at higher current density amplitude. Further studies at a constant current density and varying the fundamental excitation frequency, shows that the ion energy asymmetry is greatly reduced with a reduction in the driving frequency. Increase in the plasma asymmetry and significant *DC* self-bias at lower driving frequency is observed to be one of the principal factors responsible for the observed asymmetry in the ion energy peaks. An investigation of *DC* self-bias and plasma potential confirm that the powered electrode energy peak corresponds to the *DC* self-bias with respect to the plasma potential, and the grounded electrode peak corresponds to the plasma potential. These results suggest that although lower frequency is good for generating the asymmetry and *DC* self-bias in the discharge, but a narrow low energy *IEDF* is only possible in very high frequency driven *CCP* systems.

## 1. Introduction

Plasma technology plays a key role in microelectronics device fabrications; particularly in the processes including plasma enhanced chemical vapor deposition (*PECVD*) and plasma etching[1,2]. In such processes, the ion energy distribution function (*IEDF*) on the wafer surface is one of the crucial parameters that drive the surface reactions. A radio frequency capacitively coupled plasma (*CCP*) system, which is a dominant plasma processing tool, mostly generates a broad bi-modal shape energy distribution in which the energy dispersion is defined by the ion transit time ($\tau_i$) and frequency of the applied *RF* field ($\omega$)[3,4]. For the etching applications, as feature sizes are shrinking, one requires a narrow *IEDF* for preventing the surface damage and to increase the selectivity across different materials[5]. In *PECVD* process, controlling the *IEDF* is vital for producing desired microstructures and film properties such as a translation from amorphous to microcrystalline structures[6].

Besides the shape of *IEDF*, an independent control of ion energy and ion flux at the substrate is highly desired. In a *CCP* discharge, one way to achieve this is to employ a second frequency either on the same electrode (dual-frequency *CCPs*) or on to the opposite electrode (2-frequency *CCPs*)[7,8]. In such configurations, the flux is mostly governed by high frequency, while the low frequency controls the ion energy. Although multiple frequencies operated *CCP* discharge remains a better choice for an independent control of energy and flux, some of the earlier publications reported coupling between 2 frequencies and therefore undesired results are observed[9,10]. Another method for achieving independent control of ion flux and energy is the Electrical Asymmetry Effect (*EAE*) as proposed by *Heil et al.*[11,12]. In this method, one can generate a *DC* self-bias by changing the phase between a fundamental and its first harmonic in a dual-frequency *CCP* system. In an experimental study by Schulze *et al.*[13], verified that using variable phase angle the average ion energy varies approximately linearly with the phase and the ion flux essentially remains constant. Using this approach, the simulation results of Donkó *et al.*[14] showed the control over the shape of *IEDF* at both powered and grounded electrode. In particular, the higher moments of the *IEDF* could be varied with a change in the phase angle[15]. However, further studies for different driving frequencies, showed reduced ability of this method to control ion energy ranges at lower driving frequencies due to the secondary electron emissions[16]. Choosing electrodes of dissimilar areas, geometrical asymmetry, and varied materials are other ways for the plasma asymmetry and *DC* self-bias generation[17-20]. A combination of the discharge voltage and driving frequency has been proposed as an alternative approach for an independent control of flux and energy[21].

In recent years, the generation of electrically asymmetry using a non-sinusoidal waveform has emerged as a promising way to overcome the above challenges[22]. In this method, the use of fundamental and higher harmonics contained in the non-sinusoidal waveform, and phase between them one could generate an asymmetric plasma response even in a geometrically symmetric *CCP*. One of the examples is multi-harmonic waveforms in which the plasma density and ion flux asymmetry at one electrode increases with the number of harmonics while the average ion energy on the other electrode remains nearly constant[23-26]. Another way to produce asymmetric plasma response is to use the temporally asymmetric waveforms such as saw-tooth like waveforms[27-29]. Such type of asymmetry is known as slope *EAE*. Using such waveforms, a vast disparity in the sheath expansion adjacent to the powered and grounded electrodes generates a strong asymmetry in the ionization rate and thus a flux asymmetry is developed. This effect was further analysed and validated experimentally using Phase Resolved Optical Emission Spectroscopy (*PROES*)[30]. It was demonstrated that the single frequency generated discharge shows symmetric excitation rate at both electrodes, whereas strong excitation rate asymmetry is observed when the number of harmonics is increased to generate a sawtooth like waveform. It was noticed that by increasing the number of harmonics in the waveform i.e. turning into an ideal sawtooth like waveform further enhances the asymmetry in the discharge[27]. Furthermore, reducing gas pressure have shown the asymmetry disappears due to the shifting of ionization in the bulk plasma.

Changing the fundamental driving frequency of the applied waveform have shown drastic effect on the discharge asymmetry. In the case of sawtooth-like voltage waveform, reducing the driving frequency from 54.24 MHz to 1.695 MHz have shown the shifting the ionization peak away from the powered electrodes[27]. The lower frequency also reduces the ionization rate strongly and thus the density decreases, and symmetric density profile is observed. The combination of above 2 effects makes the discharge asymmetric as driving frequency is reduced. Similar results were observed in the case of saw-tooth like current waveform[29], where a strong asymmetry is observed at 13.56 MHz when compared to 54.24 MHz driving frequency. In addition to the discharge asymmetry, variation in the driving frequency have shown multiple ionization beams that are extended up to the opposite sheath and shown to modify the instantaneous sheath edge positions lead to the excitation of higher harmonics[29]. The frequency of the sheath modulation is observed to increase with reducing frequency and therefore claimed to be one of the parameters for driving the enhanced plasma density in the discharge at lower fundamental driving frequency of the saw-tooth like waveform. On the other hand, higher beam energy is observed at higher driving frequency, which was due to the increased sheath velocity.

Above studies mostly focused on the electrical asymmetry and the generation of *DC* self-bias in the discharge excited by non-sinusoidal waveforms. On the other hand, there exist very few studies of *IEDF* in such discharge, particularly at a very high frequency (VHF). One of the simulation studies performed by Schüngel *et al.*[31] demonstrated that by exciting plasma using a tailored waveform consist of 5 harmonics could generate a control over the shape of *IEDF*. In particular, a single peak at low/medium energies within the IEDF could be generated and controlled by adjusting the parameters of the applied voltage waveform. We follow previous studies with the investigation of *IEDF* in the *CCP* discharge using an ideal saw-tooth like current waveform at a very high frequency. In our previous work[29], we focused on the electric field non-linearity and higher harmonics generation in a symmetric *CCP*. In the present study, we extended our previous study to investigate the effect of current density amplitude on the *IEDF* at the powered and grounded electrode in a symmetric *CCP* discharge excited by sawtooth like current waveform at a fundamental driving frequency of 27.12 MHz. Additionally, we also investigate the effect of changing the fundamental driving frequency, from 13.56 MHz to 54.24 MHz, on the shape of *IEDF*. Our simulation results predict ion energy peaks corresponds to the *DC* self-bias and plasma potential. A narrow *IEDF* is observed at a higher fundamental driving frequency.

This research article is organized as follows. The simulation technique that is based on the particle-in-cell/ Monte Carlo collision (PIC/MCC) methods and simulation parameters are described in the section 2. The physical understanding and explanation of the simulation results are presented in the section 3. In section 4, the summary and conclusion are given.

## 2. Simulation Technique and Parameters

The simulation technique is based on a Particle-in-Cell/Monte Carlo collision (*PIC/MCC*) methods. For the present study, we have used a well-tested and benchmarked 1D3V, electrostatic, self-consistent, *PIC* code[32, 33]. The code has been used extensively in previous studies and details can be found in the literatures[34-46]. As an input, a current waveform is chosen given by the following mathematical expression:

$$J_{rf}(t) = \pm J_0 \sum_{k=1}^{N} \frac{1}{k} \sin(k\omega_{RF} t) \text{ -------- (1)}$$

In the above equation, the positive and negative signs resemble to "sawtooth-down" and "sawtooth-up" waveforms respectively, $J_o$ is the amplitude of current density and $\omega_{rf}$ is the fundamental angular radio-frequency (*RF*).

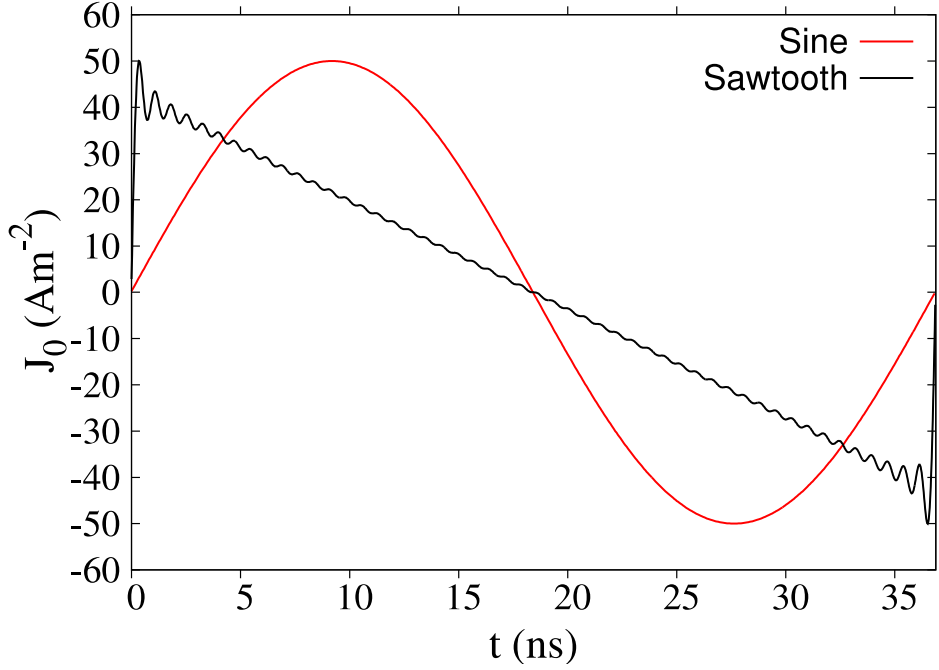

**Figure 1:** Sinusoidal and sawtooth current waveform at 27.12 MHz fundamental driving frequency and current density amplitude of 50 A/m$^2$.

An ideal sawtooth waveform is desirable to simulate the asymmetry in the discharge and therefore the waveform should consist of higher number of harmonics. In our case, we have chosen *N=50* harmonics which are distributed in such a way to provide required peak-to-peak current density amplitude for sawtooth current waveform. A profile of the sawtooth current waveform against sinusoidal waveform is shown in figure 1 for a fundamental driving frequency of 27.12 MHz. The choice of current waveform is arbitrary as the purpose of this research work is to investigate the *IEDF* for sawtooth waveform and not to replicate or comparison with any specific experiments. One of the advantages for conducting simulation for current waveforms is the validation of the analytical models prediction that is available for the constant current conditions[47,48]. In order to compare with the experiments, though exceptional, a current source can be obtained by implementing an *RF* amplifier or generator with a very high output impedance in comparison to the plasma impedance[49,50]. In this case, one can neglect the additional impedance of the plasma source and consequently current becomes the controlling parameter.

The simulation is performed in argon gas at a fixed gas pressure of 5 mTorr, which is kept constant in the discharge region throughout the simulation. The discharge gap is 6 cm. The sawtooth current waveform is applied on the electrode at 0 cm and the other electrode at 6 cm is grounded. The plasma chemistry considered in the simulation includes several particle-particle reactions such as ion-neutral (elastic, inelastic and charge exchange) and electron-neutral (elastic, inelastic and ionization). The creation of 2 metastable states, Ar$^*$, and Ar$^{**}$, is also considered. These two lumped excited states of Ar, *i.e.*, Ar$^*$ (3p54s), 11.6 eV, and Ar$^{**}$ (3p54p), 13.1 eV, in uniform neutral argon gas background is considered with charged particles viz. electrons and ions in the simulation. Important processes like multi-step ionization, metastable pooling, partial de-excitation, super elastic collisions and further de-excitation with their corresponding cross-sections are taken from well-tested source and implemented into the simulation[36, 41, 51].

The stability and accuracy criterion of PIC is achieved by selecting appropriate grid size ($\Delta x$) and time step size ($\Delta t$) that can resolve the Debye length ($\lambda_{de}$) and the electron plasma frequency respectively ($f_{pe}$)[33]. It is assumed that both the electrodes have infinite dimension and are planar, equal in size i.e. symmetric and parallel to each other. We have also considered that electrodes are perfectly absorbing and for the sake of simplicity secondary electron emission is ignored. The neutral gas is uniformly distributed with a fixed temperature of 300° K throughout the simulation. The temperature of ions is the same as the neutral gas temperature. The number of particles per cell taken here is 100 for all cases. All set of simulations run for more than 5000 *RF* cycles to achieve the steady state profile.

## 3. Results & Discussions

We first investigate the plasma density in the discharge system by changing the current density at a fundamental driving frequency of 27.12 MHz. Figure 2 (a) shows the time averaged plasma (ion) density profile within the discharge system for different current density amplitudes ranging from 10 A/m$^2$ to 125 A/m$^2$. Corresponding time averaged electron impact ionization rate, $e + Ar \rightarrow 2e + Ar^+$, is plotted in figure 2 (b). For the current set of operating parameters, it is clear that the plasma density at the centre of the discharge is increasing from $5.5\times10^{14}$ m$^{-3}$ at 10 A/m$^2$ to $1.7\times10^{16}$ m$^{-3}$ at 125 A/m$^2$. On the other hand, the effective electron temperature at the centre of the discharge is observed to decrease from ~2.7 eV at 10 A/m$^2$ to ~2 eV at 125 A/m$^2$ i.e. approximately 25% drop is noticed. As displayed in figure 2 (b), the ionization rate at the centre of the discharge increases, approximately 20 times, with a rise in current density *i.e.* from $3.0\times10^{19}$ m$^{-3}$s$^{-1}$ at 10 A/m$^2$ to $6.0\times10^{20}$ m$^{-3}$s$^{-1}$ at 125 A/m$^2$. Furthermore, due to low gas pressure and longer mean free path the ionization is mainly occur in the plasma bulk. An increase in the ionization rate and plasma density is mostly associated with the enhanced electron heating at higher current densities. In the low-pressure operating regime, the electron heating is mostly stochastic and occur near to the time-modulated sheath edge. As expressed by the hard wall model[1,52-54], the stochastic heating is proportional to the current density and therefore one will expect higher values of stochastic heating at higher current density amplitude. The simulation predicts an increase in the stochastic heating from ~250 W/m$^3$ at 10 A/m$^2$ to ~5000 W/m$^3$ at 125 A/m$^2$ *i.e.* ~ 20 times increase is observed. Subsequently, the energy gained by the electrons through interaction with the oscillating sheath dissipates into various channels through non-linear interactions including the ionization as shown in figure 2 (b), which is observed to be the highest at 125 A/m$^2$.

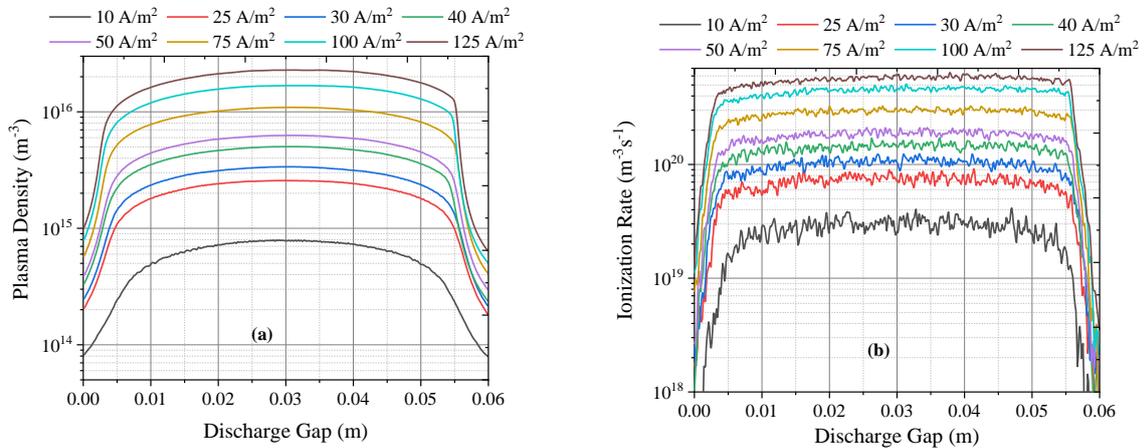

Figure 2. The profile of (a) time average plasma density and (b) time average ionization rate for different current densities from 10 A/m² to 125 A/m² at 27.12 MHz driving frequency.

Further analysis of the plasma density profile shows the generation of discharge asymmetry as the current density increases. Corresponding sheath width, estimated as where the electron sheath edge is at maximum distance from the electrode and the quasi-neutrality condition breaks down, is nearly the same (approximately 5.5 mm) at both electrodes for a current density of 10 A/m². However, as current density amplitude increases to 125 A/m² the plasma density profile becomes highly asymmetric. The sheath width at 125 A/m² is smaller, ~2.8 mm, near to the powered electrode and larger, ~4.6 mm, near to the grounded electrode. At higher current density, the asymmetry appears due to the fact that the *DC* self-bias is generated at the powered electrode because of slope asymmetry in the applied sawtooth current waveform[44]. This is shown in figure 3, where the time-average plasma potential profile is plotted in 3 (a) and *DC* self-bias at the powered electrode along with the plasma potential is plotted as a function of discharge current density amplitude. Figure 3 (a) shows that the potential is nearly the same at both powered and grounded electrode for 10 A/m². However, as current density increase, positive *DC* self-bias appears at the powered electrode. At 10 A/m² current density the *DC* self is extremely low (~1.5 V) and the plasma potential is ~29 V. As shown in figure 3 (b), both *DC* self-bias and plasma potential scales with current density amplitude. The difference between plasma potential and *DC* self-bias is also increasing with the current density amplitude.

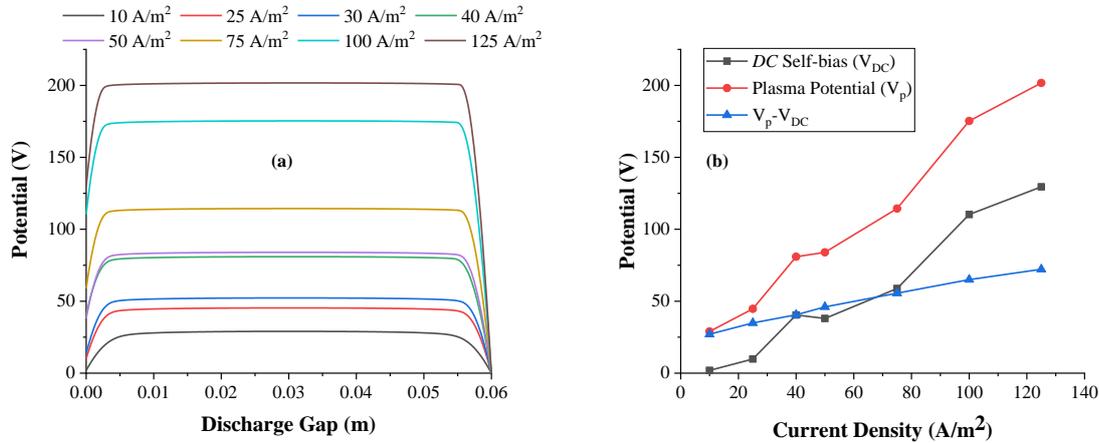

Figure 3. (a) Time-average plasma potential profile and (b) *DC* self-bias at the powered electrode and plasma potential versus current densities from 10 A/m² to 125 A/m² at 27.12 MHz fundamental frequency of sawtooth waveform.

Figure 4 (a) and 4 (b) shows the normalized *IEDF* at the powered electrode (PE) and grounded electrode (GE) respectively for different current densities from 10 A/m² to 125 A/m². The fundamental driving frequency is 27.12 MHz. The energy of the different peaks observed in the *IEDF* is also labelled. A clear asymmetry is observed in the ion energy at the powered and grounded electrode. As shown in figure 4 (a) and 4 (b), at 10 A/m², the average ion energy is lowest (~26 eV at PE and ~29 eV at GE) and single energy peaks are observed. As the amplitude of current density increases the mean ion energy increases at both PE and GE. However, the ion energy is consistently higher at the GE when compared to the PE. The asymmetry between the energy peaks at the PE and GE continues to grow with an increase in current density amplitude. At higher current densities, such as 100 A/m² and 125 A/m², bi-modal energy peaks are observed with a large energy gap between PE and GE. The difference in the energy peak is attributed to the generation of *DC* self-bias at the PE. As shown in figure 3 (b), both plasma potential and *DC* self-bias increase with a rise in current density amplitude. For the present case, the average energy of the PE peaks is consistent with the difference between plasma potential and *DC* self-bias ($V_p - V_{DC}$). On the other hand, the average energy of GE energy peaks corresponds to the plasma potential $V_p$.

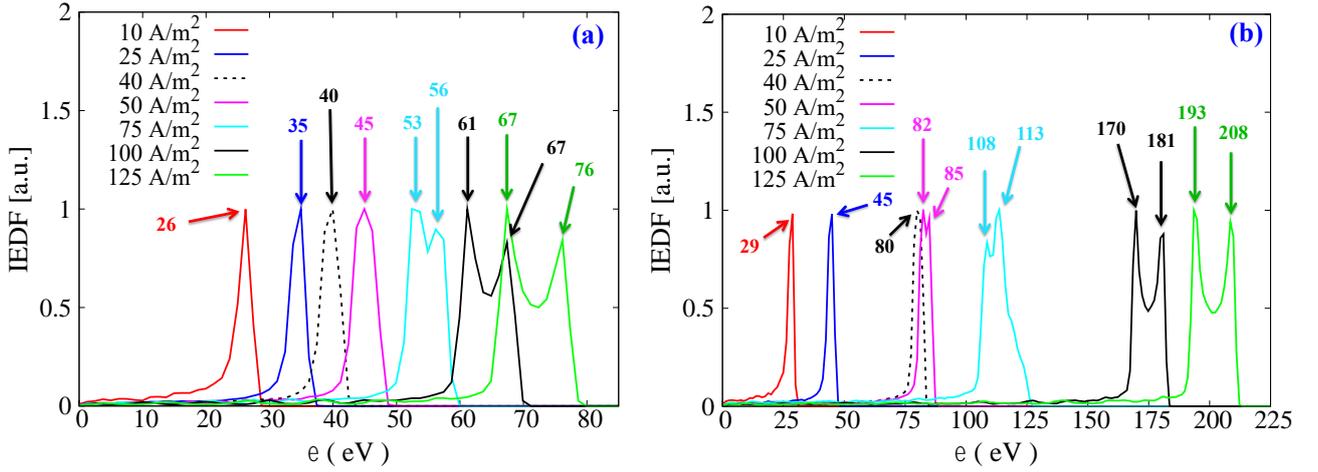

Figure 4. *IEDF* at the a) powered electrode and b) grounded electrode for current density amplitudes from 10 A/m² to 125 A/m². The fundamental excitation frequency is 27.12 MHz.

An increase in the average ion energy versus current density amplitude is attributed to a change in the sheath width and time averaged sheath voltage. This is due to the fact that ion responds to time average potential. As shown earlier, the plasma is nearly symmetric and sheath width is maximum (~5.5 mm) at 10 A/m². A typical ion mean free path for the pressure regime operating here i.e. at 5 mTorr is nearly 0.6 cm. A larger sheath width is responsible for the enhanced ion-neutral collisions. Furthermore, the time averaged potential (figure 3 (b)) at both PE and GE is lowest at 10 A/m² and therefore the average ion energy arriving at the electrode decreases. As the current density amplitude increase, the time averaged potential increase and sheath width decrease, which will reduce the ion-neutral collisions within the sheath and thus lead to higher average ion energy. On the other hand, the bi-modal behaviour of *IEDF* at higher current densities corresponds to the ion transit time ($\tau_i$) through *RF* sheath. The ion transit time is defined as $\tau_i = 3s_m/v_i \propto V_s^{1/4}/n_s^{1/2}$ where $v_i$, $s_m$, $V_s$, and $n_s$ is the ion velocity in sheath, sheath width, averaged voltage drop across sheath and density at the sheath edge respectively. For a long ion transit time $\tau_i \gg \tau_{rf}$ where $\tau_{rf}$ is the RF period defined as $\tau_{rf} = 2\pi/\omega_{rf}$. However, for the short ion transit time the condition is $\tau_i \ll \tau_{rf}$[1]. So, the ion energy distribution is broad for short ion transit time case and highly peaked for the long ion transit time case. For a constant driving frequency/*RF* period, the energy separation *ΔE* between 2 peaks of the bio-modal distribution is proportional to the sheath voltage and inversely proportional to the sheath width i.e. $\Delta E \propto \sqrt{<V_{sh}>}/<s>^3$, where $<V_{sh}>$ is the time-average sheath voltage and $<s>$ is the time-average sheath width. As observed in present work, the sheath width decreases, and averaged sheath voltage increases with a rise in current density amplitude. Therefore, bi-modal peaks with large energy separation are observed at higher current density amplitudes.

The plasma asymmetry is greatly affected by the fundamental driving frequency due to a change in the rf period. Therefore, the role of the driving frequency is crucial in determining the shape of *IEDF*. We investigate this by changing the fundamental driving frequency at a constant current density amplitude. Figure 5 shows the time-average plasma potential, 5 (a), and *IEDF*, 5 (b) at the powered electrode, 5 (c) at the grounded electrode for 3 different driving frequencies: 13.56 MHz, 27.12 MHz and 54.24 MHz at the current density amplitude of 50 A/m$^2$. As shown in figure 5 (a), the plasma is highly asymmetric at lowest driving frequency i.e. at 13.56 MHz and strong *DC* self-bias is observed at the powered electrode. As the driving frequency increase to 27.12 MHz, the plasma asymmetry and *DC* self-bias decreases. At 54.24 MHz fundamental driving frequency, the plasma profile is nearly symmetric, and the *DC* self-bias is nearly zero at the powered electrode. This behaviour is mainly attributed to the ionization asymmetry as discussed in our previous article[44].

Further analysis of *IEDF* shape presented in figure 5 (b) and 5 (c), depicts bi-modal distribution at both PE and GE for 13.56 MHz fundamental driving frequency. The average energy of the PE peak is ~106 eV that corresponds to the *DC* self-bias with respect to the plasma potential, whereas the average energy of the GE peak is ~317 eV that coincide with the time averaged plasma potential. As the driving frequency increase to 27.12 MHz, bi-modal peaks turned into single energy peaks at both PE and GE, and the ion energy asymmetry also decreases i.e., PE peak is observed at 45 eV and averaged ion energy at the GE is at 83.5 eV. Finally, at 54.24 MHz, narrow nearly symmetric energy peaks are observed at PE and GE with ~36 eV and ~ 32 eV ion energy respectively. The conversion into single energy peaks as driving frequency increase is attributed to a drop in the time averaged potential as shown in figure 5 (a). Furthermore, as driving frequency decrease the rf period increase and thus broadening of the energy ion peak is anticipated i.e. $\Delta E \propto <V_{sh}>/\omega$. At 13.56 MHz, strong asymmetry in the ion energy at the PE and GE is due to the generation of large *DC* self-bias (~210 V). Whereas the *DC* self-bias is lowest (~4 V) at 54.24 MHz fundamental driving frequency, which lead to a reduced asymmetry in the average ion energy.

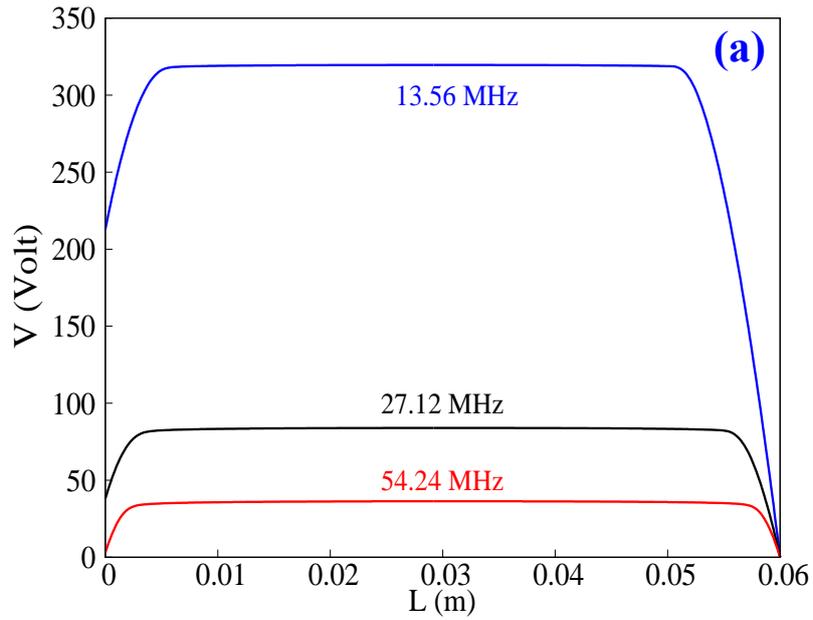
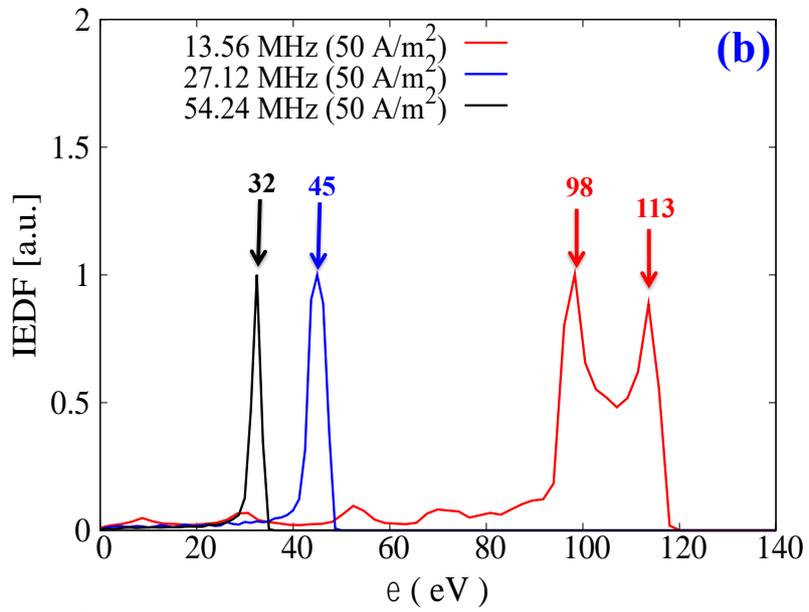
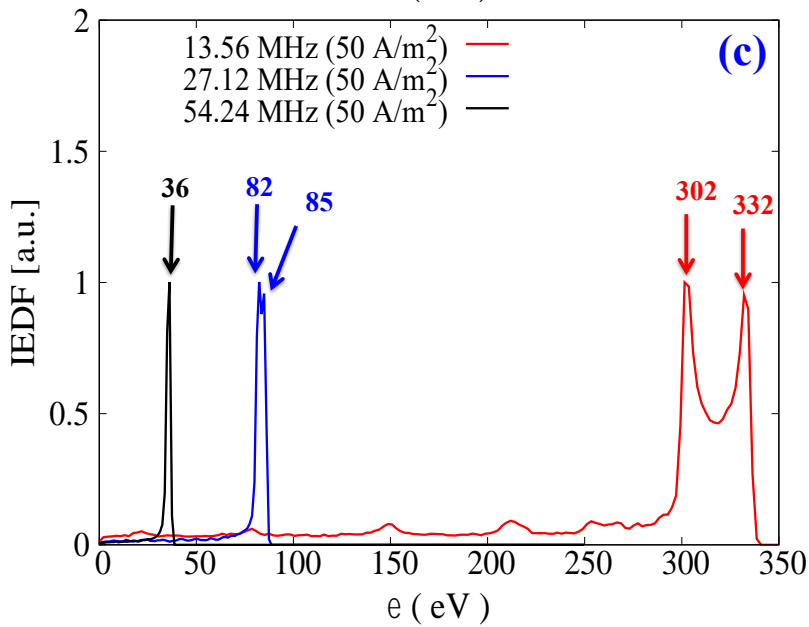

Figure 5. (a) Time-average plasma potential profile, (b) *IEDF* at the powered electrode and (c) *IEDF* at the grounded electrode for 13.56 MHz, 27.12 MHz and 54.24 MHz fundamental frequencies of a sawtooth waveform at 50 A/m$^2$ current density amplitude.

## 4. Summary of the work and conclusions

The study of plasma density, *DC* self-bias/plasma potential and *IEDF* is performed using *PIC* simulation is a geometrically symmetric *CCP* discharge excited by a sawtooth like current waveform at a fixed gas pressure of 5 mTorr. At a fundamental driving frequency of 27.12 MHz, the plasma density observed to increase with the current density amplitude. This is due to fact that the stochastic heating increases with current density amplitude and hence the ionization rate also increases. The time average plasma potential observed to increase along with the formation of the *DC* self-bias that scales linearly with the current density amplitude. At lower current density amplitude, the asymmetry is lower and the *IEDF* at the powered and grounded electrode shows low energy and single energy peak. As current density amplitude increase, bi-modal *IEDF* is observed with a large energy separation between the powered and grounded electrode i.e. the asymmetry is higher. A close observation of the average energy peaks suggests that this corresponds to the formation of *DC* self-bias and therefore the multiple average energies of the peak relates with the *DC* self-bias and plasma potential. Changing the fundamental driving frequency of the sawtooth current waveform reveals that the *IEDF* could be converted into a narrow single energy peak, which is attribute to the symmetric behaviour of the discharge at higher driving frequency. For the simulation results it is concluded that although lower fundamental driving frequency of the saw-tooth waveform is capable for generating discharge asymmetry and *DC* self-bias, but a narrow *IEDF* is only possible at a higher driving frequency due to a reduced rf period and decrease in the averaged sheath voltage.

**Acknowledgments**

Dr A Kuley is supported by the Board of Research in Nuclear Sciences (BRNS Sanctioned No. 39/14/05/2018-BRNS), Science and Engineering Research Board EMEQ program (SERB Sanctioned No. EEQ/2017/000164), National Supercomputing Mission (NSM) (Ref No: DST/NSM/R&D_HPC_Applications/2021/04) and Infosys Foundation Young Investigator grant.

**Data Availability**

The data that support the findings of this study are available from the corresponding author upon reasonable request.